\documentstyle[twocolumn,floats,aps,epsf]{revtex}

\def\ltsim{\vbox {\hbox{\lower 0.9\baselineskip \hbox{$<$}} \break
         \hbox{\lower 0.2\baselineskip \hbox{$\sim$}} } }

\def\gtsim{\vbox {\hbox{\lower 0.9\baselineskip \hbox{$>$}} \break
         \hbox{\lower 0.2\baselineskip \hbox{$\sim$}} } }

\begin{document}

\twocolumn[\hsize\textwidth\columnwidth\hsize\csname@twocolumnfalse%
\endcsname

\title{Details of disorder matter in 2D d-wave superconductors} 
\author{W. A. Atkinson$^1$, P. J. Hirschfeld$^1$, A. H.
MacDonald$^2$, and K. Ziegler$^3$}
\address{$^1$Department of Physics, University of Florida, PO Box 118440,
Gainesville FL 32611}
\address{$^2$Department of Physics, Indiana University,
Swain Hall W.\ 117, Bloomington IN 47405}
\address{$^3$ Institut f\"ur Physik, Universit\"at Augsburg, 86135 Augsburg,
Germany}
\date{\today}
\maketitle
\draft
\begin{abstract}
We demonstrate that discrepancies between predicted low-energy
quasiparticle properties in disordered 2D d-wave superconductors
occur because of the unanticipated importance of disorder model
details and normal-state particle-hole symmetry.
This conclusion follows from numerically exact evaluations of 
the quasiparticle density-of-states predicted by the 
Bogoliubov-deGennes (BdG) mean field equations for both binary alloy 
and random site energy disorder models.  For the realistic case,
which is best described by a binary alloy model  
without particle-hole symmetry, we predict density-of-states
suppression below an energy scale which appears to be correlated
with the corresponding single-impurity resonance.
\end{abstract}


]
\narrowtext

In metals, the many different analytic and numerical techniques
used to study disorder and electron localization have led
to a satisfyingly  consistent understanding.  The same cannot
be said for the theory of disorder and quasiparticle localization in
high $T_c$ superconductors, which are widely believed to be correctly
modelled as 2D systems with $d$-wave pairing.  As pointed out some
years ago by Nersesyan, Tsvelik and Wenger\cite{Nersesyan}, the
standard perturbative approximation for the self-energy, the
self-consistent T-matrix approximation (SCTMA), breaks down in 2D
$d$-wave superconductors as $|E| \rightarrow 0$, even in the dilute
impurity limit.  In response, a variety of non-perturbative approaches
have been applied\cite{Nersesyan,Ziegler,Fisher,PepinLee,Zirnbauer},
and have yielded apparently contradictory results.  The purpose of
this Letter is to demonstrate, by exact numerical calculation, that
these discrepancies occur for the most part because, in contrast with
the metallic case, details of the disorder model are qualitatively
important.  Moreover, for a physically important class of disorder
models, the seemingly innocent assumption of particle-hole symmetric
normal state bands leads to non-generic results.

This work focuses on the quasiparticle density-of-states (DOS)
$\rho(E)$, which is strongly affected by even small impurity
concentrations.  In pure materials, $d$-wave superconductivity is
characterized by a gapless density of states $\rho(E) \sim |E|$ for
$|E| < \Delta_0$ ($\Delta_0$ is the $d$-wave gap amplitude). The
SCTMA\cite{Gorkov,SR,Hirschfeld}, predicts a finite DOS for disordered
materials at $E=0$.  Non-perturbative approaches beyond the SCTMA have
variously predicted that $\rho(E)$ vanishes according to
universal\cite{Nersesyan,Fisher} power laws, that $\rho(E)$
diverges as $|E| \rightarrow 0$\cite{PepinLee,Zirnbauer}, and that there is a
rigorous lower bound on $\rho(E)$\cite{Ziegler}.

At first sight it seems impossible that these results could be
mutually reconciled.  We will undertake to show here, however, that
most can be understood within a single framework, and that they differ
primarily because of details in the treatment of disorder.  It is
natural to assume that such dramatic discrepancies
for what a priori appear to be only slightly different physical models
arise because the $d$-wave system is critical.  We compare models by
solving the Bogoliubov-de Gennes equations numerically on large
finite-size lattices.  The model parameters we vary include disorder
type (see below), disorder strength, normal state particle-hole
symmetry, self-consistent renormalization of the local order parameter
by disorder, and the so-called ``Dirac cone anisotropy"
$v_F/v_\Delta$, where $v_F$ is the Fermi velocity and $v_\Delta$ is
the velocity of quasiparticles transverse to the nodes.  We find that
binary alloy and random site energy disorder models differ
qualitatively.  For strong scatterers, in particular, random site
disorder models cannot describe the enhancement in the low-energy DOS
predicted by P\'epin and Lee\cite{PepinLee}, which we reproduce here.
In addition, some of the present authors have recently shown that
self-consistent treatment of the order parameter cannot be neglected
in general.\cite{atkinson} We suggest that an appropriate model for
disorder in the cuprates must involve a binary alloy treatment of
strongly scattering impurities and self-consistency, and predict in
this case a power-law $\rho(E)\sim E^\alpha$ with disorder-dependent
$\alpha$ below an energy scale set by the single impurity
resonance\cite{atkinson}.

{\it Method.}
We consider a mean-field Bogoliubov-de Gennes Hamiltonian 
for electrons hopping on a tight-binding square lattice
with nearest neighbor hopping matrix element $t$, and bond
mean field order parameter $\Delta_{ij}$,
\begin{eqnarray}
\cal{H} &=& - t \sum_{\langle i,j\rangle} \sum_{\sigma}
c^\dagger_{i\sigma} c_{j\sigma} - \sum_{i,\sigma} [\mu - U_i]
c^\dagger_{i\sigma} c_{i\sigma} \nonumber \\ &-& \sum_{\langle
i,j\rangle} \{ \Delta_{ij} c^\dagger_{i\uparrow}
c^\dagger_{j\downarrow} + h.c. \},
\label{eq:ham}
\end{eqnarray}
where $U_i$ is the impurity potential on site $i$.  Energies will be
measured in units of the hopping amplitude $t$ and lengths in units of
the lattice constant.  We consider both random site energy models, in
which the $U_i$ are chosen randomly from a distribution $P(U)$, and
binary alloy models, in which $U_i$ takes the value $U_0$ on a
fraction $n_i$ of the sites and is zero elsewhere.  The filling is
chosen to stabilize a pure $d$-wave ground state in the absence of
disorder, with homogeneous order parameter $\Delta_k = \Delta_0 [
\cos(k_x) - \cos(k_y) ]$, where $ \Delta_0 = \frac{1}{2}\sum_\pm (
\Delta_{i\,i\pm x} - \Delta_{i\,i\pm y} ) $.  In the tight-binding
model, $v_F/v_\Delta = 2t/\Delta_0$.

For the $d$-wave system, Eq. (\ref{eq:ham}) has been used to calculate
DOS,\cite{Xiang}superfluid density\cite{Franz,Randeria}, and $T_c$
supression\cite{Franz} numerically.  It is widely assumed\cite{Xiang}
that local fluctuations in $\Delta_{ij}$ produced by the impurity
potential do not affect the DOS qualitatively.  Recent numerical
work\cite{atkinson,Randeria} has suggested otherwise, and we therefore
make a distinction between self-consistent (SC) solutions of the BdG
equations, where $\Delta_{ij} \equiv V_{ij} \langle c_{j\downarrow}
c_{i\uparrow}\rangle$ with nearest neighbor pairing interaction
$V_{ij}$, and non-self-consistent (NSC) calculations, where
$\Delta_{ij}$ has the homogeneous $d$-wave form.  With $\Delta_{ij}$
determined, we can calculate the DOS from $\rho(E) = L^{-2}
\sum_\alpha \delta(E-E_\alpha)$, where $E_\alpha$ are the eigenvalues
of ${\cal H}$, for samples of size $L\times L$.  Our numerical
calculations were performed on systems with $L \leq 45$, and real
periodic and antiperiodic boundary conditions.

{\it Random site energy vs. binary alloy models.}  We begin by
discussing the NSC models used in the vast majority of earlier work.
In field-theoretical approaches it is common to assume a Gaussian
distribution $P(U)= (\sigma\sqrt{2\pi})^{-1}\exp (-U^2/2\sigma^2)$ for
the disorder potential at each site. For technical reasons it is more
convenient to consider a uniform ``box'' distribution $P(U)=1/(2W)$,
$|U|<W$.  We have checked that results with the box distribution are
similar to the Gaussian, with the mapping $W\simeq\sqrt{3}\sigma$.
Disorder is thus characterized by a single parameter, in contrast to
the binary alloy model, where chemical impurities or vacancies are
characterized not only by their individual scattering strength $U_0$,
but also by their concentration $n_i$.  In the normal metal, a
correspondence between $W$ and $(n_i,U_0)$ can always be found such
that the random site energy and binary alloy models yield similar
results.  This is no longer true in the superconducting state, because
the frequency dependence of the superconducting Green's function can
lead to midgap resonances\cite{Salkola}, found only in the binary
alloy case and observed in experiment\cite{PanDavis,Yazdani}.

\begin{figure}[tb]
\begin{center}
\leavevmode
\epsfxsize \columnwidth
\epsffile{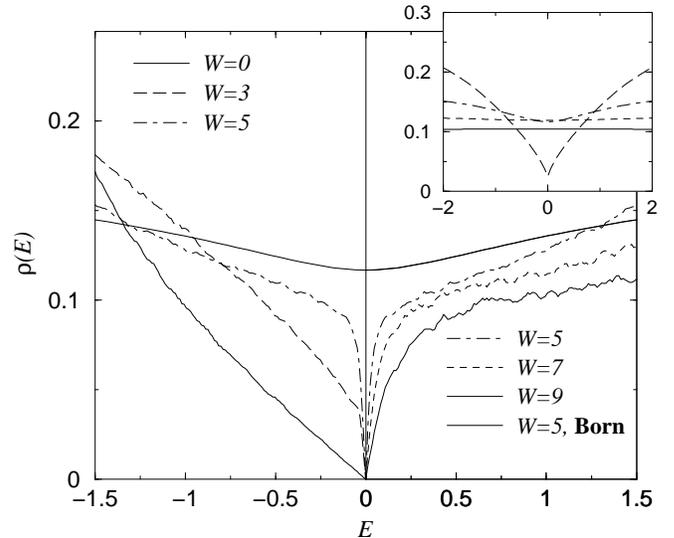}
\caption{Density of states for box distributed disorder. Main figure:
Exact NSC solution of the BdG equations for different $W$.  For
comparison, the Born approximation for $W=5$ is also shown.  Inset:
Born approximation for different $W$.  Line types refer to same values
of $W$ as in main figure ($W=0$ not shown).  $\Delta_0 = 2$
and $\mu=1.2$.}
\label{fig:box1}
\end{center}
\end{figure}

\begin{figure}[tb]
\begin{center}
\leavevmode
\epsfxsize \columnwidth
\epsffile{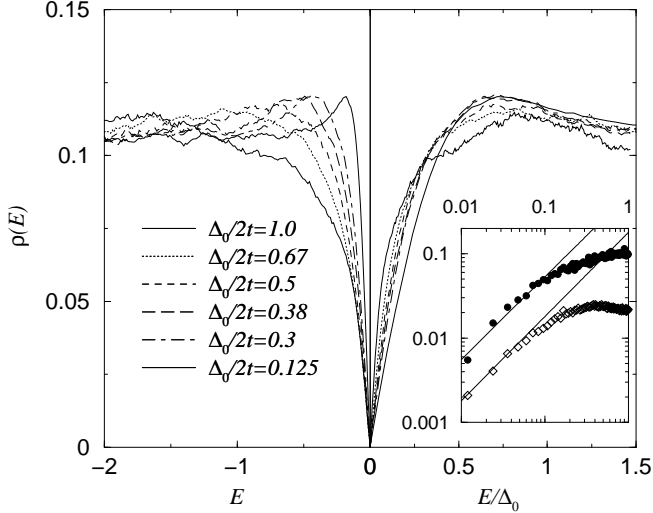}
\caption{Scaling of the DOS with $\Delta_0$ for $\mu = 1.2$ and $W=9$.
Inset: Logarithmic plots for $\Delta_0/2t = 1$ (circles) and
$\Delta_0/2t = 0.3$ (diamonds; data rescaled for clarity).  Lines
are guides to the eye indicating linearity.}
\label{fig:box2}
\end{center}
\end{figure}

Figure \ref{fig:box1} shows the DOS of a $d$-wave superconductor at
low energies in the presence of box disorder at different $W$.  When
$W$ is small, the self-consistent Born limit reproduces
the exact calculation for the box distribution quantitatively.
For larger $W$, the exact calculation shows the formation of a
``pseudogap'' over an energy interval $|E| < E_1$, where $E_1$ grows
rapidly with $W$.  The physics of the pseudogap is clearly not
captured by the Born limit approximation for the box distribution,
which predicts a finite residual DOS $\rho_{\mathrm mf}$.  In
Fig.~\ref{fig:box2}, we study the pseudogap in more detail.  With
$v_F=v_\Delta$, and large disorder, we can identify a second, much
smaller energy scale $E_2$ over which $\rho(E) \sim |E|$.  This regime
disappears quickly, however, as $v_\Delta$ is decreased, and we
contrast this behaviour with the relatively slow scaling of $E_1$ with
$v_\Delta$.  Earlier field theoretical studies\cite{Fisher} made
predictions for a linear DOS over an energy scale $\sim
1/\rho_{\mathrm mf} \xi_L^2$.  The rapid scaling of $E_2$ in
Fig.~\ref{fig:box2} is consistent with the predicted exponential
dependence of $\xi_L$ on $v_F/v_\Delta$.  It is also tempting to make
the connection between the pseudogap edge $E_1$ and the much larger
predicted scale for weak localization corrections to the
DOS\cite{PALee,Zirnbauer}.  In this case, $E_1$ is expected to scale
as $\Delta_0 \rho_{\mathrm mf}$ for small disorder, and it is clear
from Fig.~\ref{fig:box2} that the scaling with $\Delta_0$ holds 
even for large disorder.

{\it Unitary limit.}  There is considerable evidence that simple
defects in the CuO$_2$ planes give rise to local scattering centers
close to the unitarity limit, indeed that simple defects in all
unconventional superconductors scatter with phase shifts close to
$\pi/2$, for reasons which are not completely understood.  For
isolated impurities, the signature of unitarity is a resonance in the
local DOS at $E= 0$; resonances close to $E=0$ have been observed in
recent scanning tunneling microscopy
experiments\cite{PanDavis,Yazdani} for both ``native defects" and Zn
atoms substituting on the planar copper sites in BaSrCaCuO-2212. A
point which has not been widely appreciated is that the value of the
impurity potential $U_0$ which produces a unitary resonance is
dependent on the band structure; for a perfectly symmetric band the
unitary limit corresponds to $U_0 \rightarrow \pm\infty$\cite{Salkola}, 
while for an asymmetric band, $U_0$ is a finite value
dependent on the degree of
asymmetry\cite{Fehrenbacher,Joynt,atkinsonII}. There is a fundamental
distinction between the two cases, with the first exhibiting perfect
particle-hole symmetry on all energy scales\cite{PepinLee}, and the
second exhibiting particle-hole symmetry on energies $|E| < \Delta_0$.
Most analytical treatments of the disorder problem do not distinguish
between the two, using $\Delta_0$ as a high energy cutoff.  We show
below that in the many-impurity problem erroneous conclusions may be
drawn as a result.

\begin{figure}[h]
\begin{center}
\leavevmode
\epsfxsize \columnwidth
\epsffile{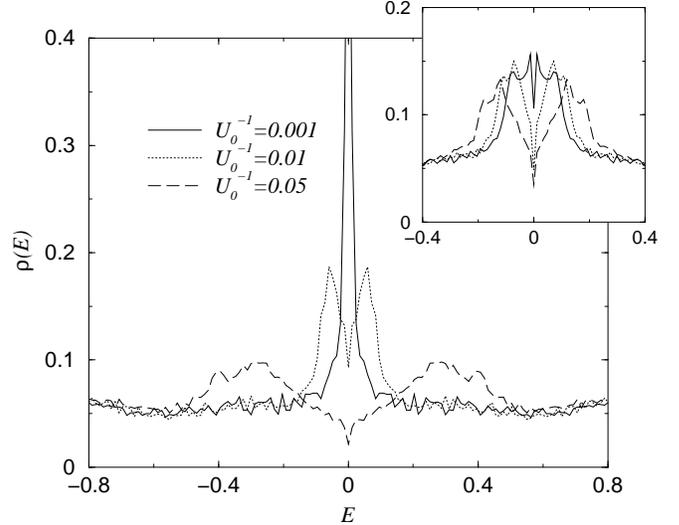}
\caption{NSC calculation of DOS near the unitary limit for $\Delta_0 =
2$.  Main figure: $\mu =0$.  Inset: $\mu = 0.2$
with $U_0^{-1} = 0.001$ (solid), 0.01 (dotted), 0.02
(dashed).}
\label{fig:unitary}
\end{center}
\end{figure}

In the unitary limit, the SCTMA predicts that a plateau forms in the
DOS over an energy interval $|E| < \gamma$.  Recent non-perturbative
calculations by P\'epin and Lee\cite{PepinLee} found that unitary
scatterers produce a divergent DOS $\rho(E) \sim
1/|E|\ln^2(|E|/\Delta_0)$ as $|E| \rightarrow 0$.  This feature was
not found in recent numerical work by some of the current
authors\cite{atkinson}, who considered only tight-binding bands with
$\mu \neq 0$.  In Fig.~\ref{fig:unitary}, we show that a divergent DOS
occurs only in models with particle-hole symmetry at all energy
scales, for $\mu=0$ and $U_0^{-1}= 0$ in our case.  The effects of
breaking particle-hole symmetry are illustrated in
Fig.~\ref{fig:unitary}.  For $\mu=0$ and $U_0^{-1} \neq 0$, the
divergent peak splits, and moves away from $E=0$ as $|U_0^{-1}|$
grows.  This is qualitatively similar to what happens in the single
impurity limit\cite{Salkola}, although the peak splitting occurs more
rapidly in the bulk disordered case.  An alternative means of breaking
particle-hole symmetry is to let $\mu\neq 0$, in which case the peak
structure rapidly disappears (Inset, Fig.~\ref{fig:unitary}).  This
differs from the single impurity limit where there always exists some
finite $U_0$ at which a zero energy divergence occurs.  A remnant of
the isolated impurity resonance remains, however, as a broad
accumulation of states at low energies.  The extra spectral weight is
not reproduced by SCTMA calculations, and it gradually vanishes as we
move further away from perfect symmetry.  Evidently the low-energy
effective action is unstable both to deviations from unitarity and to
deviations from full band particle-hole symmetry.

\begin{figure}[tb]
\begin{center}
\leavevmode
\epsfxsize \columnwidth
\epsffile{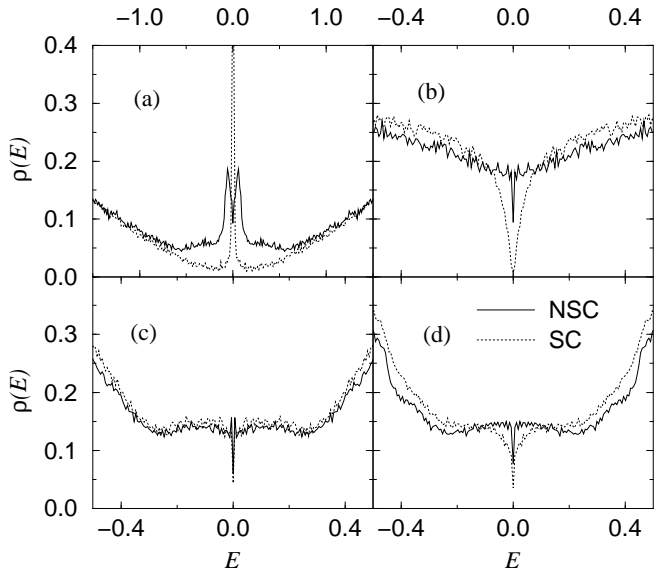}
\caption{Exact solutions of BdG equations with (SC) and without (NSC)
order parameter self-consistency.  (a) Binary alloy, symmetric band,
near unitarity; $\Delta_0=2$, $U_0=100$, $n_i=0.05$, $\mu=0$.  (b) Box
distribution; $\Delta_0=0.39$, $W=2$, $\mu=1.2$ (c) Realistic binary
alloy model; $\Delta_0=0.39$, $U_0 = 5$, $n_i=0.04$, $\mu=0.6$.
(d) As in (c) but with $\mu=1.2$.}
\label{fig:SC}
\end{center}
\end{figure}

{\it Local order parameter suppression.} In the above discussion, we
have concentrated on the artificial case in which the order parameter
was not allowed to vary spatially, with the purpose of understanding a
set of disparate theoretical results obtained under this assumption.
As discussed in Ref.~\cite{atkinson}, allowing $\Delta_{ij}$ to
respond self-consistently to the impurity potential introduces a new
source of scattering in the off-diagonal channel, which ultimately
leads to a suppression of the DOS at low energy.  This effect
highlights the complex nature of multi-impurity scattering resonances,
and is opposite to the expectations of the naive ``Swiss cheese''
picture, in which the impurity simply produces a small region of
normal metal.  In Fig.~\ref{fig:SC} we show the effect of
self-consistency in the unitary limit.  In the case of a symmetric
band, self-consistency moves the resonance towards the Fermi level
(consistent with what is seen in the single impurity
case\cite{Shnirman,atkinsonII}) and also suppresses the overall low
energy DOS.  This kind of DOS suppression is also seen in SC solutions
with box distributed disorder [Fig.~\ref{fig:SC}(b)].  In the binary
alloy model, however, it is possible to make an empirical connection
between the energy scale for DOS suppression and the energy of the
single impurity resonance.  The energy scales are not equal, but
clearly scale together\cite{atkinson}.  One interesting implication is
that for scatterers sufficiently close to the unitary limit, the
pseudogap will be unobservable, as illustrated in Fig.~\ref{fig:SC}(c),
but that the same impurity doped into a different host may
cause the pseudogap to open [Fig.~\ref{fig:SC}(d)].  We speculate that
this may be the case with Zn, which is clearly a unitary scatterer in
BSCCO\cite{PanDavis}, but appears to produce a depression in the DOS
at small energies in LSCO, as seen in recent specific heat
experiments\cite{Lin}.


{\it Conclusions.}  We have shown that a number of different
approaches to the $d$-wave disorder problem which produce apparently
contradictory results can be understood within a single framework when
the appropriate symmetries of the Hamiltonian, and in particular of
the particular realization of disorder, are accounted for.  The most
important question we hope to settle here is which of the preceding
results, if any, are of relevance to experiment.  We remind the reader
that experiments which probe the DOS most directly are consistent with
the existence of a constant DOS at the Fermi
energy\cite{Scalapinoreview}.  We have shown that a true constant DOS
{\it cannot} be understood in 2D $d$-wave superconductors with any of
the disorder models discussed here\cite{footnote1}, and we suggest
several possible reasons for the discrepancy.  The first possibility
is that experiments are unable to access the pseudogap regime in the
optimally doped materials because of native near-unitarity
defects\cite{PanDavis}.  In this case, the only effects of weak
localization on the DOS would be the weak nonmonotonicity shown in
Fig.~\ref{fig:SC}(c),(d), which is reminiscent of effects seen in the
specific heat.\cite{Kapitulnik} The second possibility is that weak
coupling to the third dimension destroys the DOS anomalies described
above; we expect on general grounds that the influence of the crossed
diagrams identified by Nersesyan et al. \cite{Nersesyan} will become
negligible in the dilute limit in 3D, and the validity of the SCTMA
will be restored.  In this context we note that almost all the
experiments indicating finite residual DOS in the cuprates have been
performed on YBCO, the most 3D of the cuprate materials.  We hope our
work will serve as an incentive to examine the low-energy properties
of disordered 2D materials like BSCCO-2212.  Finally, we mention the
possibility that many-body effects beyond the BCS approximation play
an important role at low energies.  There is some speculation that
this might be the case in the underdoped cuprates, and the formulation
of a relevant theory is an interesting but difficult problem which
must be left to future research.

{\it Note added}: In the  final stages of preparation of this
manuscript, we received a preprint from Zhu et al.\cite{Ting}
in which similar results for the unitarity limit of the symmetric
band were obtained.

{\it Acknowledgements}

This work is supported by NSF grants DMR-9974396, DMR-9714055, and
INT-9815833.  The authors would like to thank M.P.A. Fisher,
D. Maslov, K.\ Muttalib, S. Vishveshwara, and A.  Yashenkin for useful
discussions.


\begin{thebibliography}{stuffstuffstuff}
\bibitem{footnote1} Although the claimed result of finite residual
density of states $\rho(0)$ by Ziegler et al. \cite{Ziegler} may still
be  correct, it is clear from this work that $\rho(E)$ cannot
resemble the SCTMA result as suggested.
\end{thebibliography}
\end{document}